\documentclass{appolb}

\usepackage{epsfig}

\begin{document}
\pagestyle{plain}
\title{Lepton Flavour Violation at the ILC%
\thanks{Presented at XXVII PHYSICS IN COLLISION - Annecy, France, 26 - 29 June 2007.}%
}
\author{P.M. Ferreira$^{1}$~\footnote{ferreira@cii.fc.ul.pt}, R.B. Guedes$^{1}$~\footnote{renato@cii.fc.ul.pt}
and R. Santos$^{1,2}$~\footnote{rsantos@cii.fc.ul.pt}
\address{$^{1}$ Centro de F\'{\i}sica Te\'orica e Computacional, Faculdade de Ci\^encias,
Universidade de Lisboa, Avenida Professor Gama Pinto, 2, 1649-003
Lisboa, Portugal \\ $^{2}$
Department of Physics, Royal Holloway,
University of London, Egham, Surrey TW20 0EX United Kingdom\\[-0.5cm]}
} \maketitle
\begin{abstract}
\noindent We explore the possibility of detecting lepton flavour
violation, now a well established experimental fact, at the
International Linear Collider. Using a model independent approach
we conclude that, given all experimental constraints available,
there is still room to detect lepton flavour violation at the ILC.
\end{abstract}
%
%
\noindent This work is based on {\it Lepton flavour violating
processes at the International Linear
Collider}$^{}$~\footnote{P.~M.~Ferreira, R.~B.~Guedes and
R.~Santos, Phys.\ Rev.\  D {\bf 75} (2007) 055015.} (please see
paper for the complete list
of references).\\
The effective operator formalism is based on the assumption that
the Standard Model (SM) is the low energy limit of a more general
theory. Such a theory would be valid at very high energies but, at
a lower energy scale $\Lambda$, we would only perceive its effects
through a set of effective operators of dimensions higher than
four.
%
%
%
%
%
The effective operators that are important for our studies fall
into two categories: those that generate lepton flavour violating
(LFV) vertices of the form $Z \,l_h\, l_l$ and $\gamma \,l_h
\,l_l$, where $l_h$ and $l_l$ are a heavy lepton and a light
lepton, respectively, and four-fermion operators, involving only
leptonic spinors.
%
%
The effect of these LFV operators can be seen in two types of
decays: a heavy lepton decaying into three light ones,
$l_h\,\rightarrow\,l\,l\,l$ and a Z boson decaying to two
different leptons, $Z\,\rightarrow\,l_h\,l_l.$
Experimentally, the upper bound on the branching ratio for $\tau$
decaying into three light leptons is of the order of $10^{-7}$ and
of order $10^{-12}$ for the $\mu.$ The bounds on the $Z$ branching
ratios are of the order $10^{-6}$ for $Z\,\rightarrow\,e\,l_h$ and
$10^{-5}$ for $Z\,\rightarrow\,\tau\,\mu.$
We concentrate on the most straightforward processes where LFV
could be detected at the ILC: $e^+e^- \rightarrow \mu^- e^+,$ $e^+
e^- \rightarrow \tau^- e^+,$ and $e^+ e^- \rightarrow \tau^-
\mu^+,$ as well as
the respective charge conjugates.\\[0.2 cm]
%
%
%
We computed the cross sections and the decay widths for these LFV
processes and used the experimental constraints to limit the
possible range of the anomalous couplings.
The range of values chosen for each of the coupling constants was
$10^{-4} \leq |a/\Lambda^2| \leq 10^{-1}\,TeV^{-2}$, where $a$
stands for a generic coupling and $\Lambda$ is in TeV. For $a
\approx 1$ the scale of new physics can be as large as 100 TeV.
This means that if the scale for LFV is much larger than 100 TeV,
it will not be probed at the ILC unless the values of the coupling
constants are unusually large. We then generate random values for
all anomalous couplings (four-fermion and $Z$ alike), and discard
those combinations of values for which the several branching
ratios we computed earlier are larger than the corresponding
experimental upper bounds. The following figure shows us the
example of the number of events occurring at the ILC for the
process $e^+\,e^-\,\rightarrow\,\tau^-\,  e^+$ in terms of the
branching ratio $BR(\tau\,\rightarrow\,lll).$\\[-0.8cm]
\begin{figure}[h]
 \begin{center}
    \epsfig{file=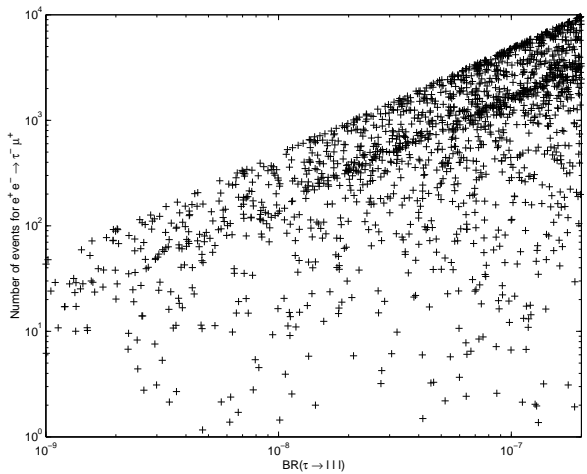,width=7.8 cm,angle=0}
   \caption{\small{Number of expected events with a center-of-mass
   energy of 1 TeV and a total luminosity of $\mathbf{1\,\, ab^{-1}}$.}}
  \end{center}
\end{figure}
\\[-0.7cm]
We have considered all planed future experiments on LFV. In the
foreseeable future, the constraints on the four-fermion $\tau$
couplings could decrease one order of magnitude. Therefore, even
in this case, the maximum number of events at the ILC would be
$\sim 1000$ and detection of LFV at the ILC would still be
possible in that case. We also studied all possible SM backgrounds
that could mask this signal. For example, for the signal process
$e^+ e^- \rightarrow \tau^+ \, e^- \rightarrow  \mu^+ \, e^- \,
\nu_{\mu} \, \bar{\nu}_{\tau}$ we have evaluated the two main
backgrounds, $e^+ e^- \rightarrow \mu^+ \, e^- \, \nu_{\mu} \,
\bar{\nu}_{e}$ and $e^+ e^- \rightarrow \tau^+ \, \tau^-
\rightarrow \mu^+ \, e^- \, \nu_{\mu} \, \bar{\nu}_{e} \,
\nu_{\tau} \, \bar{\nu}_{\tau}$. With appropriate cuts on angular
variables and transverse momentum, we were able to obtain a clear
signal.
\end{document}